\documentclass[9pt,twocolumn,twoside]{pnas-new}

\templatetype{pnasinvited} 

\usepackage{graphicx}
\usepackage{epsfig}
\usepackage{epstopdf}
\usepackage{subfigure,color}
\usepackage{dcolumn}
\usepackage{bm}
\usepackage{lettrine} 

\def\_#1{{\bf #1}}

\title{Dynamic meta-atoms} 

\author[1,*]{Grigorii~Ptitcyn}
\author[1,*]{Mohammad~S.~Mirmoosa}
\author[1]{Sergei~A.~Tretyakov} 

\affil[1]{Department of Electronics and Nanoengineering, Aalto University, P.O.~Box 15500, FI-00076 Aalto, Finland.*These authors contributed equally: Grigorii Ptitcyn, Mohammad S. Mirmoosa. *email: grigorii.ptitcyn@aalto.fi and mohammad.mirmoosa@aalto.fi} 

\authorcontributions{} 
\authordeclaration{} 
\equalauthors{} 
\correspondingauthor{} 


\begin{abstract}
Interaction of electromagnetic radiation with time-variant objects is a fundamental problem whose study involves foundational principles of classical electrodynamics. Such study is a necessary preliminary step for delineating the novel research field of linear time-varying metamaterials and metasurfaces. A closer look to the literature, however, reveals that this crucial step has not been addressed and important simplifying assumptions have been made. Before proceeding to studies of linear time-varying metamaterials and metasurfaces with their effective parameters, we need to rigorously describe the electric and magnetic responses of a temporally-modulated meta-atom. Here, we introduce a theoretical model which describes a time-variant meta-atom and its interaction with incident electromagnetic waves in time domain. The developed general approach is specialized for a dipole emitter/scatterer loaded with a time-varying reactive element. We confirm the validity of the theoretical model with full-wave simulations. Our study is of major significance also in the area of nanophotonics and nano-optics because the optical properties of all-dielectric and plasmonic nanoparticles can be varied in time in order to achieve intriguing scattering phenomena. 
\end{abstract} 

\dates{} 
\doi{}

\begin{document} 
\maketitle
\ifthenelse{\boolean{shortarticle}}{\ifthenelse{\boolean{singlecolumn}}{\abscontentformatted}{\abscontent}}{} 

%%%%%%%%%%%%%%%%%%%%%%%%%%%%%%%%%%%%%%%%%%%%%%%%%%%%%%%%%%%%%%%%%%%%%%%%%%%%%%%%%%%%%%%%%%%%%%%%%%%%%%%%

\begin{figure}[t!]\centering
\includegraphics[width=0.48\textwidth]{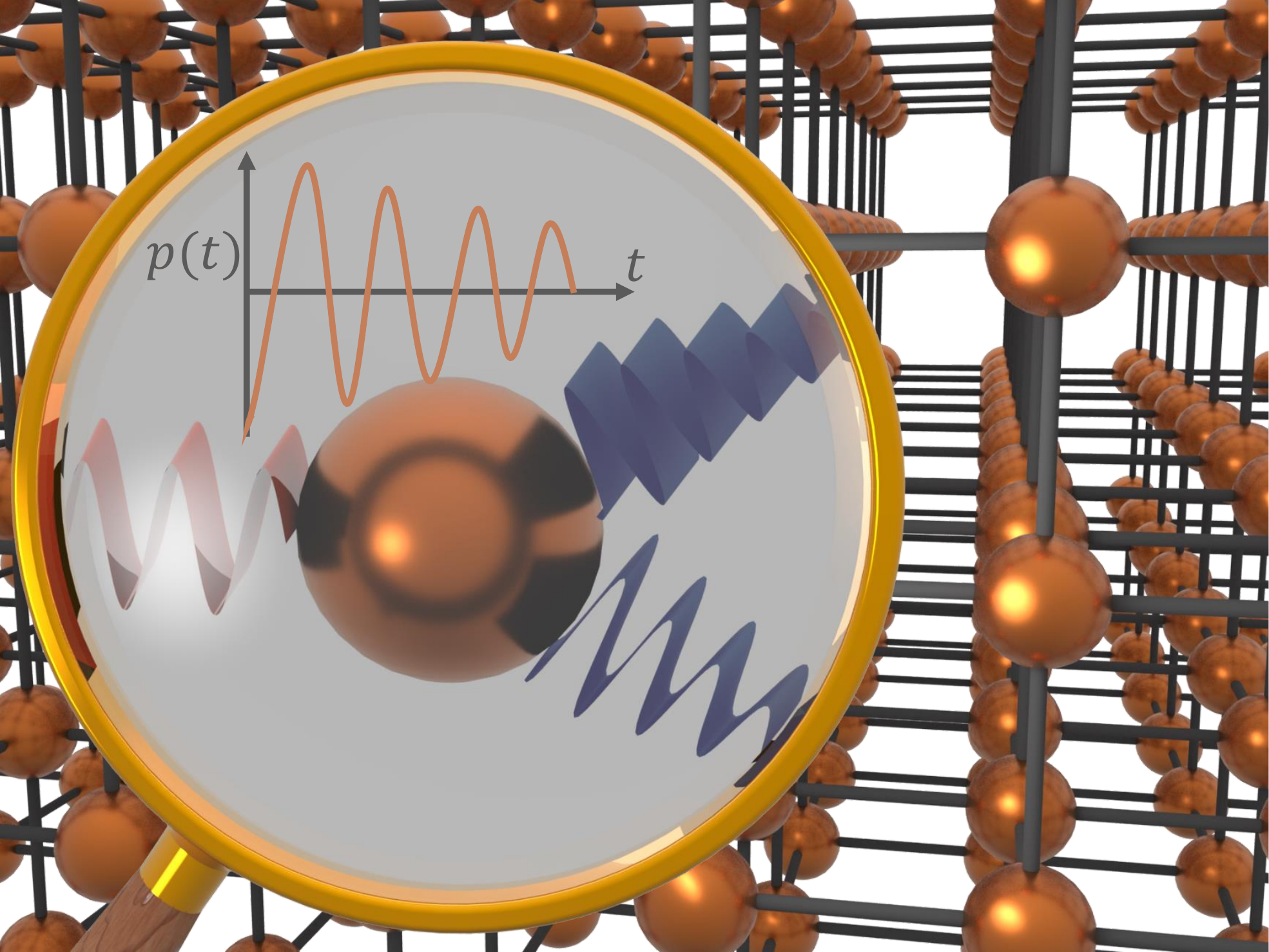}
\caption{{\bf Schematic view of a metamaterial formed by  time-varying meta-atoms}. Here, the meta-atom is assumed  to exhibit only electric response determined by its electric dipole moment ${\_p}(t)$. The excitation and temporal variations of the dipole moment can be controlled and engineered using time modulation of the meta-atom.} 
\label{fig_metameta}
\end{figure}

\lettrine[lines=3]{T}{}~he physical properties of metamaterials~\cite{Wegener} and metasurfaces~\cite{Tretyakovphysics} can be engineered by proper selection of materials, sizes, shapes, and mutual orientations of meta-atoms. These custom-designed artificial materials and surfaces have recently enabled significant advances in  controlling  electromagnetic radiation and realizing new physical phenomena, especially in electromagnetics and optics. Time-varying external modulation is an important degree of freedom which can potentially open new interesting possibilities in microwave techniques, optoelectronics, photonics, and other fields (see e.g.~Refs.~\cite{Fleuryletter,Fleurynature,Aluphoton,Fanletter,Kamal,Sounasnature,Wang,Krishnaswamy,Estep,Cullen,Tzuang_Fan_Lipson,Shmidt_Lipson,Phare_Lipson,Lira_Fan_Lipson}). Known studies consider phenomena in \emph{materials with time-varying parameters} (usually assuming that material permittivity~\cite{Fannature,Alu_Sounas,Fleury_sounas_Alu,Cassedy_Oliner,Taravati_Caloz,Shi_Han_Fan} or conductivity~\cite{Dinc_Alu} depends on time due to some external force), but do not explore the possibilities of shaping material response by time modulations of constituent  \emph{meta-atoms}. We expect that, by using time modulations at the microscopic level of meta-atoms, radically new approaches  can be found to shape the effective properties of materials formed by these meta-atoms, see illustration in Fig.~\ref{fig_metameta}. Furthermore, such novel approaches can be applied for rather general manipulations and optimizations of resonant and non-resonant small objects (antennas, plasmonic and dielectric nanoparticles, quantum dots, etc.)

\begin{figure*}[t!]\centering
\includegraphics[height=5.8cm]{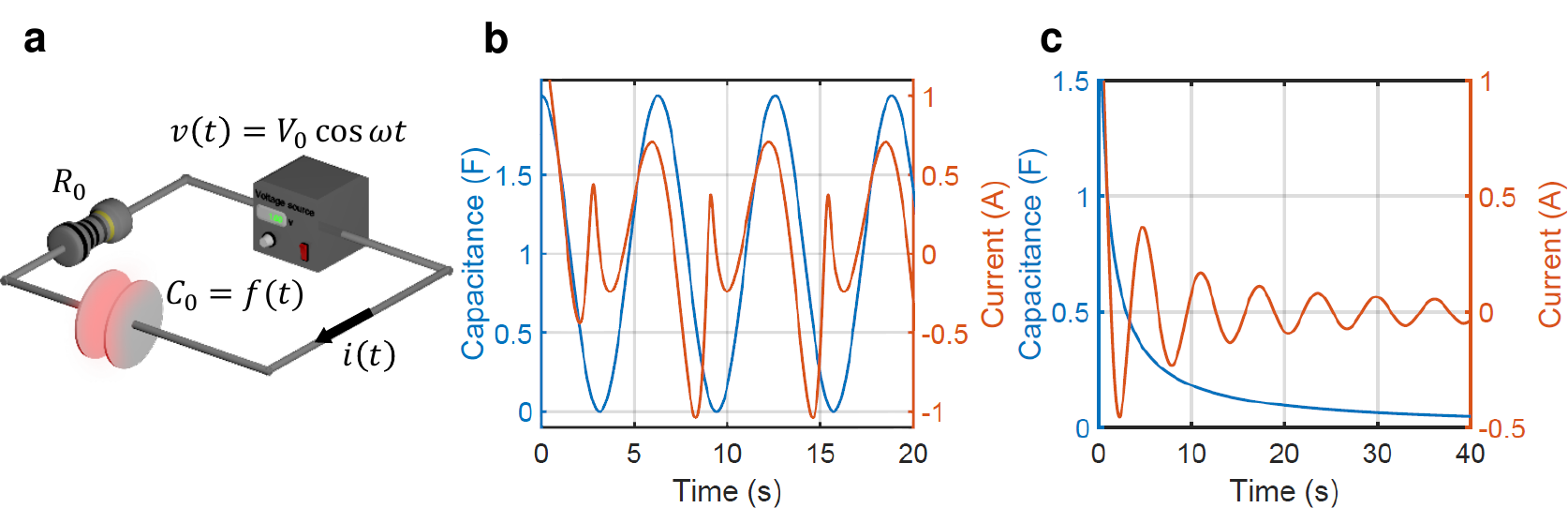}
\caption{{\bf Engineering of electric current in a series $RC$ circuit.} {\bf a--b}, Time-varying capacitance and the corresponding electric current flowing through the electric circuit described by {\bf{c}}. {\bf c}, Schematic view of the $RC$ circuit. In this example, $R_0=1~\Omega$, $V_0=1~{\rm{V}}$, and $\omega=1~{\rm{rad/s}}$. The resistance and the capacitance of the circuit are connected to a time-harmonic voltage source $v(t)=V_0\cos(\omega t)$.} 
\label{fig_RC_circuit}
\end{figure*}

However, there are two fundamental issues which need to be addressed before this novel research area can be properly developed and explored. The known approaches assume that the effective material or surface parameters depend on time, while the fundamental constituent of metamaterials is the meta-atom. By imposing some arbitrary time-dependence to a property such as permittivity, one has to make dramatic simplifying assumptions, namely assume that the  modulation is slow~\cite{Kang_Russell,Liu_Shadrivov,Hadad_Alu,Gupta2018} and usually assume that the material response is instantaneous. Obviously, a material cannot polarize instantaneously in response to an applied field. Polarization at a given moment of time $t_0$ depends on the field and electric susceptibility at previous times, $t<t_0$, such that for linear and stationary media the polarization is a convolution of susceptibility and the electric field. Instantaneous response corresponds to a susceptibility in form of Dirac delta function. Using time-domain models with time-varying permittivity [$\_D(t)=\epsilon(t)\_E(t)$] one assumes instantaneous response, therefore such approach does not allow studies of dispersive time-modulated media. Using frequency-domain models with the permittivity dependent both on the frequency and time (due to external modulation) one assumes that the modulation is very slow at the scale of all polarization processes in the material. 
It appears that only for lossless plasma described by the Drude model there are known models which take into account frequency dispersion of material response under external time modulations of material parameters (the plasma frequency, in this case)~\cite{Nerukh}.    
In addition, studies of time-modulated materials, metamaterials and metasurfaces, have been limited mostly to time-harmonic modulations~\cite{Hadad_Alu,Liu_Shadrivov,Gupta2018} or to sharp, step-wise~\cite{sun2016optical} changes in the material parameters. 

The intriguing question is:  Can we find such (non-harmonic, arbitrary) time modulations of meta-atoms which will enable new phenomena which are not achievable with conventional time-harmonic pumping? In other words, in addition to studying phenomena in media which are time modulated in a particular (usually time-harmonic) way, one can possibly create and optimize desired effects by choosing an appropriate time-modulation function. It appears that this question has not been posed and addressed till today, and there is a clear need for fundamental investigations in this direction. This paper makes a  step in this direction by studying arbitrary time modulations of dipolar  meta-atoms. Here, we consider one single meta-atom which is modulated by an external force and create a theoretical model for its description in time domain. Using this model, we discuss several new application scenarios where the time dependence of induced polarization and meta-atom scattering response can be fully engineered  by an external time modulation. 

%%%%%%%%%%%%%%%%%%%%%%%%%%%%%%%%%%%%%%%%%%%%%%%%%%%%%%%%%%%%%%%%%%%%%%%%%%%%%%%%%%%%%%%%%%%%%%%%%%%%%%%%%%%%

\section*{Circuit analogy}

An analogy between an $RLC$ circuit and a meta-atom is well known in the context of time-harmonic excitation and response~\cite{Tretyakov1,Stas}. The dissipation of energy including radiation is modeled by $R$ (resistance), and the reactive elements $L$ (inductance) and $C$ (capacitance) are associated with storing of magnetic and electric energies, respectively. The electric current flowing through the resistor of the equivalent circuit is related to the induced dipole moment of the meta-atom. The meta-atom can be excited either by external electric fields or by an internal source (voltage or current source). This analogy motivates us to contemplate the scenario where one of the reactive elements of the circuit is temporally modulated. Let us examine an $RC$ circuit, shown in Fig.~\ref{fig_RC_circuit}, and draw a possible inference assuming that the capacitance is time-varying $C_0=f(t)$. Generalizing the conventional theory of parametric circuits with time-harmonic pumping, we suppose that function $f(t)$ is arbitrary. We expect that selecting the modulation function will make it possible to engineer the current flowing through the circuit in rather general ways, for any  given external electromotive force $v(t)$.  
%Explicitly, if the lumped resistance and the capacitance are constant (time-invariant) and we connect them in series to an ideal time-harmonic voltage source, the electric current is deterministically time-harmonic in the steady-state regime. Hence, we cannot engineer the electric current such that we have our desired temporal function for the electric current. However, theory of circuits shows that utilizing time-varying capacitance changes the story. 
Indeed, for this circuit, the electric current can be expressed as $i(t)=\exp\Big(-\int a(t)dt\Big)\Big(\int b(t)\exp\Big(\int a(t)dt\Big)dt+A\Big)$, where $a(t)$ is a function of $C_0=f(t)$ and of the time derivative of $f(t)$. Parameter $b(t)$, in addition, depends on the voltage source and on  the time derivative of the external voltage (see Supplementary Information). We can expect that properly choosing the time-varying capacitance will result in a desired temporal function for the electric current. Here, we emphasize that the exciting voltage $v(t)$ can be  time-harmonic, still allowing for arbitrary shaping of the time dependence of the current. Figure~\ref{fig_RC_circuit} shows two examples of current time dependencies for different temporal modulations of  capacitance. The electric current flowing in the circuit is not time-harmonic and, in principle, it can be an arbitrary function. For example, in paper \cite{us} it is shown that properly modulating inductive load, reflections from the load can be completely eliminated. 

In the time-harmonic scenario and in the frequency domain, interactions of small meta-atoms with electromagnetic waves  can be studied using an antenna model of an electrically small dipole~\cite{Tretyakov1} and its equivalent circuit. On the basis of this model, the electric current carried by the dipole is determined in terms of the frequency-dependent input impedance ($Z_{\rm{in}}$) of the dipole and  the impedance of the load connected to the dipole center ($Z_{\rm{Load}}$) as $I=\_E{_{\rm{inc}}}\cdot \_l/(Z_{\rm{in}}+Z_{\rm{Load}})$, where $\_E{_{\rm{inc}}}$ is the complex amplitude of the incident electric field impinging on the Hertzian dipole (meta-atom) and $l$ is the effective length of the  dipole. The numerator $\_E{_{\rm{inc}}}\cdot \_l$ models  an ideal voltage source: the electromotive force induced by the incident wave. Close to the resonance of small meta-atoms, we can model the dipole as a series $RLC$ circuit. Now, it can be readily perceived what can happen if the reactive load of the  dipole is time-varying, since the system is similar to what we discussed above: An $RLC$ circuit with a time-varying reactive element. In consequence, it can be expected that one can engineer the electric current of the meta-atom in the desired fashion or, in other words, we are free to manipulate the response of the meta-atom to the incident electromagnetic wave. Note that the electric current is proportional to the time derivative of the electric dipole moment ($\mathbf{p}$) which is related to the incident wave field ($\mathbf{E}_{\rm{inc}}$) by the electric polarizability ($\alpha_{\rm{ee}}$). In this sense, the electric current induced in the meta-atom defines the effective permittivity of a medium, because that is determined by the polarizablities of meta-atoms. 

However, as we will show next, solutions for arbitrary time-modulated meta-atoms cannot be in general found using the  Fourier transform of the frequency-domain solutions. The main reason is that the conventional expression for the radiation resistance of the dipole is valid only for time-harmonic fields and gives only time-averaged radiated power. In the literature, the problem of calculation of instantaneous power radiated from dipoles was addressed in~\cite{Schantz,Kaiser,Valagiannopoulos_Alu,Vandenbosch1,Vandenbosch2,Papas_Pulsed,GSSmith}. However, the attention was mostly on instantaneous reactive power around dipole. The known results do not self-consistently consider the source of radiation and therefore do not allow writing time-domain dynamic equations for currents on radiating and time-modulated dipole scatterers. More general models are necessary, which we develop here.

%%%%%%%%%%%%%%%%%%%%%%%%%%%%%%%%%%%%%%%%%%%%%%%%%%%%%%%%%%%%%%%%%%%%%%%%%%%%%%%%%%%%%%%%%%%%%%%%%%%%%%%%%%%

\section*{Instantaneous power balance}

Conventionally, the problem of small or Hertzian dipole under time-harmonic excitation is treated through the method of phasors which allows to express the fields in the far and near zones in a simple way~\cite{Jackson,Balanis}. Using these expressions, it is possible to describe the radiation via fundamental parameters of radiators and scatterers such as the radiated power or radiation resistance on the basis of antenna theory~\cite{Balanis}. Such characterization utilizes quantities that are averaged in time over a period, making this approach suitable only for time-harmonic current densities or for slowly modulated meta-atoms. In this study we treat a problem of arbitrary temporal modulations which means that the electric current is not necessarily close to time-harmonic at the time scale of several periods. Therefore, we must find alternative methods providing us with the ability to describe the interaction of the dipole with the electromagnetic radiation in time domain, without the use of time averaging in calculations of radiated power and radiation resistance. We will approach this problem using the principle of conservation of energy.

In the circuit theory, there is the law of instantaneous power balance which is basically an expression of the conservation of energy at each moment of time. According to this law, the summation of all instantaneous powers, related to resistance, capacitance, inductance and the sources of the circuit, must be zero ($\sum_k P_k(t)=0$). The Kirchhoff current and voltage laws can be derived from the instantaneous power balance. For small dipoles, according to the analogy with the $RLC$ circuit, we can also use the instantaneous power balance equation. The external instantaneous power $P_{\rm{in}}(t)$ supplied to the dipole splits into three parts: the radiated power $P_{\rm rad}(t)$, the reactive power $P_{\rm reac}(t)$ and the dissipated power $P_{\rm diss}(t)$ (if the dipole contains some resistance; from the meta-atom point of view it is the absorbed power due to the inherent losses in the particle materials). Therefore, we can write the following equation which represents the power balance for the dipole:
\begin{equation}
P_{\rm in}(t)+P_{\rm reac}(t)+P_{\rm rad}(t)+P_{\rm diss}(t)=0.
\label{eq_powers}
\end{equation}
As mentioned above, in the case of time-harmonic excitation and time-invariant elements, we use the  values of power that are averaged over a period. Average of the reactive power is always zero and it does not contribute to the power balance in the frequency domain. However, for time-modulated meta-atoms the instantaneous reactive power even in the case of a time-harmonic source is not zero and,  since we treat the problem in the time domain, the reactive power cannot be dropped out. This issue is important because the reactive energy represents the stored energy near the dipole in the form of electric and magnetic field energies.

Regarding the radiated power, it is defined as the surface integral of the Poynting vector over a closed surface including the  dipole~\cite{Griffith,Balanis}. According to the antenna theory terminology, the radiation of the dipole is described through the concept of radiation resistance $R_{\rm{r}}=80\pi^2(l/\lambda)^2$, where $l$ is the effective dipole length and $\lambda$ denotes the wavelength~\cite{Balanis}. This result stems from the time-harmonic calculation of the time-averaged radiated power and the time-average of the square of the instantaneous electric current carried by the dipole. In other words, 
\begin{equation}R_{\rm rad}=\langle P_{\rm rad}\rangle/\langle i(t)^2\rangle ,
\label{Rrad}
\end{equation}
where the brackets denote time averaging over one period. Obviously, this model is applicable only for time-harmonic currents or for modulations which are very slow as compared with the carrier frequency. For arbitrary time modulations, this description of radiation is not suitable and we must find a general expression for the radiation resistance or the instantaneous radiated power such that the expression for the radiated power is valid for any moment of time. To this end, it is not enough to calculate the fields created by the  dipole and subsequently compute the instantaneous Poynting vector. Later, we numerically show that such computation can give rise to wrong interpretations even in the case of time-harmonic excitation and no time modulation.

Let us consider the general scenario of radiation from currents, i.e.~the electric current density can be an arbitrary function of time and space (bound inside a small volume, so that the dipole model is valid). Is it possible to derive a general expression for the radiated power? We show that it is possible if we consider the fields inside the volume containing all radiating currents and determine how much energy and momentum from moving charges are transferred to electromagnetic fields at any moment of time. In this way, we can rigorously determine the instantaneous radiated power. For this purpose, we express the Poynting theorem in form
\begin{equation}
-\int_{V_{\rm{c}}}\_{J}\cdot\_E\,dv=P+\frac{dU}{dt}.
\label{eq_energy_transfer}
\end{equation}
Here, $\_E$ is the electric field generated by $\_J$ and $V_{\rm{c}}$ is the volume of the radiating system. The above expression states, in fact, the conservation of energy: The time rate of change of energy which flows out through the boundary surface of the volume ($P$) plus the time rate of change of energy within the volume ($dU/dt$) is equal to the time rate at which the energy is transferred to the electromagnetic fields (the negative of work done by the fields on $\_J$). 

%According to  Eq.~(\ref{eq_energy_transfer}), there are two different paths for achieving the instantaneous radiated power $P_{\rm{rad}}(t)=P(t)$. The first path is through the right side of Eq.~(\ref{eq_energy_transfer}), and the second path is through the left side of that equation. We must certainly get the same result for the radiated power. Here, as mentioned above, we choose the second path which is related to the electric current density and the electric field. In Eq.~(\ref{eq_energy_transfer}), we must integrate over the current volume $V_{\rm{c}}$, consisting of charges, and 

Let us consider radiation from a known distribution of current density, bound to a small volume. Knowing the current density, we can find the exact electric field $\_E$ distribution inside the  radiating volume. The electric field is expressed via the gradient of the scalar potential $\phi$ and the vector potential $\_A$ (i.e.~$\_E=-\nabla\phi-\partial\_A/\partial t$), in which
\begin{equation}
\begin{split}
&\phi(\_r ,t)=\frac{1}{4\pi\epsilon_0}\int\frac{\rho\big(\displaystyle\_r^\prime,t-\frac{1}{c}|\_r-\_r^\prime|\big)}{|\_r-\_r^\prime|}d\_r^\prime,\cr
&\_A(\_r ,t)=\frac{\mu_0}{4\pi}\int\frac{\_J\big(\displaystyle\_r^\prime, t-\frac{1}{c}|\_r-\_r^\prime|\big)}{|\_r-\_r^\prime|}d\_r^\prime.
\end{split}
\label{eq_potentials}
\end{equation}
Here, $\rho$ denotes the charge density, and $c$ is the speed of light. Within the nonrelativistic approximation and assuming that the radiating current volume is electrically small, expansion of the charge and current densities as power series of $|\_r-\_r^\prime|/c$ allows us to find the dipolar radiated power. These expansions are explained in detail in Supplementary Information of this paper. Making those expansions, some algebraic manipulations result in
\begin{equation}
\begin{split}
&-\int_{V_{\rm{c}}}\_J\cdot\_E\,d\_r\approx\cr&-\frac{\mu_0}{6\pi c}\_{\dot{p}}(t)\cdot\_{\dot{\ddot{p}}}(t)+
\frac{d}{dt}\Bigg[\frac{1}{8\pi\epsilon_0}\int_{V_{\rm{c}}}\frac{\rho(\_r,t)\rho(\_r^\prime,t)}{|\_r-\_r^\prime|}d\_r\,d\_r^\prime\cr
&+\frac{\mu_0}{16\pi}\int_{V_{\rm{c}}}\frac{\partial\rho(\_r,t)}{\partial t}\frac{\partial\rho(\_r^\prime,t)}{\partial t}|\_r-\_r^\prime|\,d\_r\,d\_r^\prime\cr
&+\frac{\mu_0}{8\pi c}\int_{V_{\rm{c}}}\frac{\_J(\_r,t)\cdot\_J(\_r^\prime,t)}{|\_r-\_r^\prime|}\,d\_r\,d\_r^\prime\Bigg],
\end{split}
\label{eq_energy_transfer_all_terms}
\end{equation} 
where $\_p=\int_{V_{\rm{c}}}\_r^\prime\rho(\_r^\prime,t)\,d\_r^\prime$ denotes the electric dipole moment (recall the assumption that the source volume is small on the scale of all relevant wavelengths).
The above equation must include the total stored electromagnetic energy $U$ and the radiated power $P$~(see Eq.~(\ref{eq_energy_transfer})). The full time derivative terms written inside brackets $[\dots ]$ are associated with the electromagnetic energy (note that the singularity seen in these terms is resolved by using the method of principal value~\cite{IsmoLindell}). Therefore, the radiated power can be written as 
\begin{equation}
P(t)=-\frac{\mu_0}{6\pi c}\_{\dot{p}}(t)\cdot\_{\dot{\ddot{p}}}(t).
\label{Power}
\end{equation}
This equation is of major significance because it reminds us the Lorentz friction force~\cite{Landau} describing the fact that radiation exerts a reaction force back on the source. Interestingly, the above expression can be written as a summation of two terms:  
\begin{equation}
\_{\dot{p}}(t)\cdot\_{\dot{\ddot{p}}}(t)=-\_{\ddot{p}}(t)\cdot\_{\ddot{p}}(t)+\frac{d}{dt}\Big(\_{\dot{p}}(t)\cdot\_{\ddot{p}}(t)\Big).
\label{eq_energy_transfer_two_terms}
\end{equation}
In the literature, the first term, $\_{\ddot{p}}(t)\cdot\_{\ddot{p}}(t)$, is considered as the dipole radiation power which propagates infinitely far~\cite{Griffith} while the second term is usually set aside because it is a full time derivative and therefore it vanishes in the time-averaged scenario. However, here, we cannot discard this term since we are interested in the  instantaneous power balance. Below, we present example full-wave simulations to confirm this important statement. Note that in the typical time-harmonic case, the first term results in the expression $\omega^4 p/2$, where  $\omega$ is the angular frequency and $p$ is the amplitude of the oscillating electric dipole moment~\cite{Griffith}, corresponding to the conventional radiation resistance \eqref{Rrad}.   

The established time-domain power balance equation can be used as the governing time-domain equation for currents on radiating and scattering time-modulated meta-atoms. Within the dipolar approximation,  we can write the complete version of the instantaneous power balance equation
\begin{equation}
\begin{split}
&-\frac{\mu_0}{6\pi c}\_{\dot{p}}(t)\cdot\_{\dot{\ddot{p}}}(t)+\frac{Q(t)}{C}i(t)+L\frac{di(t)}{dt}i(t)+\frac{Q(t)}{C_{\rm{load}}(t)}i(t)\cr
&+\frac{d}{dt}\Bigg(L_{\rm{load}}(t)i(t)\Bigg)i(t)-\mathcal{E}(t)i(t)=0
\end{split}
\label{eq_inst_power_balence_plus_modulated_load}
\end{equation}
in terms of the effective capacitance $C$ and inductance $L$ of the meta-atom. Here, $C_{\rm{load}}(t)$ and $L_{\rm{load}}(t)$ are the time-varying load capacitance and inductance, respectively. The load is in series connection with the reactive elements of the antenna, $C$ and $L$. Charge at one arm of the dipole is denoted as $Q(t)$. The electromotive force $\mathcal{E}(t)$ can be created by either external electric field of waves illuminating the meta-atom or by an external source connected to the dipole arms. Solutions of this equation give accurate relations for the time dependence of induced electric dipole moment of a radiating meta-atom which is arbitrarily modulated in time. 

%%%%%%%%%%%%%%%%%%%%%%%%%%%%%%%%%%%%%%%%%%%%%%%%%%%%%%%%%%%%%%%%%%%%%%%%%%%%%%%%%%%%%%%%%%%%%%%%%%%%%
\begin{figure*}[t!]\centering
\includegraphics[height=6cm]{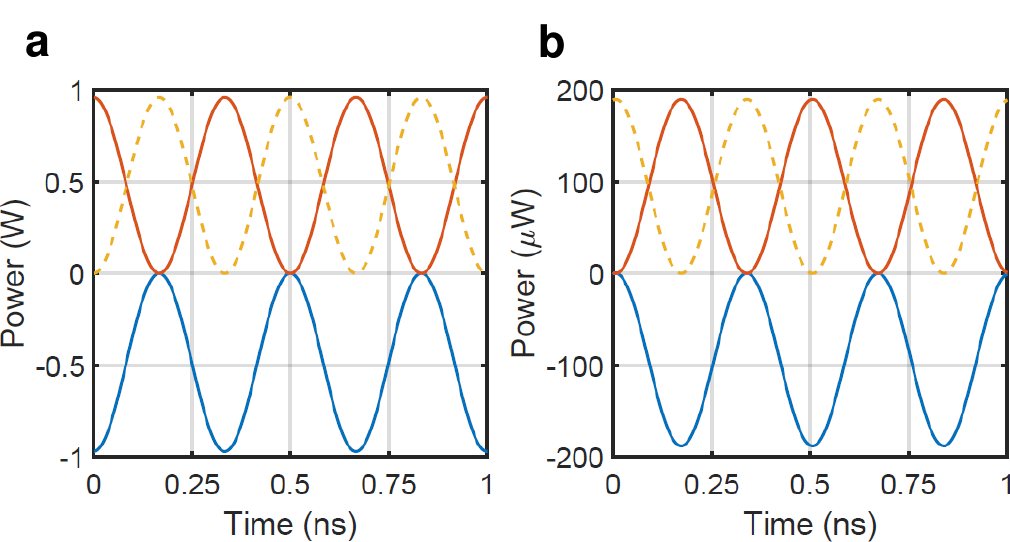}
\caption{{\bf Instantaneous power balance for transmitting regime regarding two different dipoles.} {\bf a}, Resonant dipole: The length of the cylindrical wire, made of perfect conductor, is $l=93.9~\rm mm$ and the radius of the wire is $r=0.3~\rm mm$. {\bf b}, Non-resonant dipole: The length of the wire is $l=15~\rm mm$ in this case, which is much smaller than the wavelength that is 200~mm. The radius of the wire remains the same $r=0.3~\rm mm$. In both {\bf a} and {\bf b}, the orange solid curve shows the instantaneous radiated power, while the blue solid curve corresponds to a sum of the reactive and supplied powers. Additionally, the dashed curve represents the  conventional interpretation of the instantaneous radiated power, i.e $P(t)=-(\mu_0/6\pi c)\_{\ddot{p}}(t)\cdot\_{\ddot{p}}(t)$, which is seen to be incompatible with the instantaneous conservation of energy. In {\bf a}, the total reactive power is zero due to the resonance, and the blue curve simply describes the power that is supplied to the resonant dipole.}
\label{fig_Pow_balance_dipoles_transmitting}
\end{figure*}

\begin{figure*}[t!]\centering 
\includegraphics[height=5.5cm]{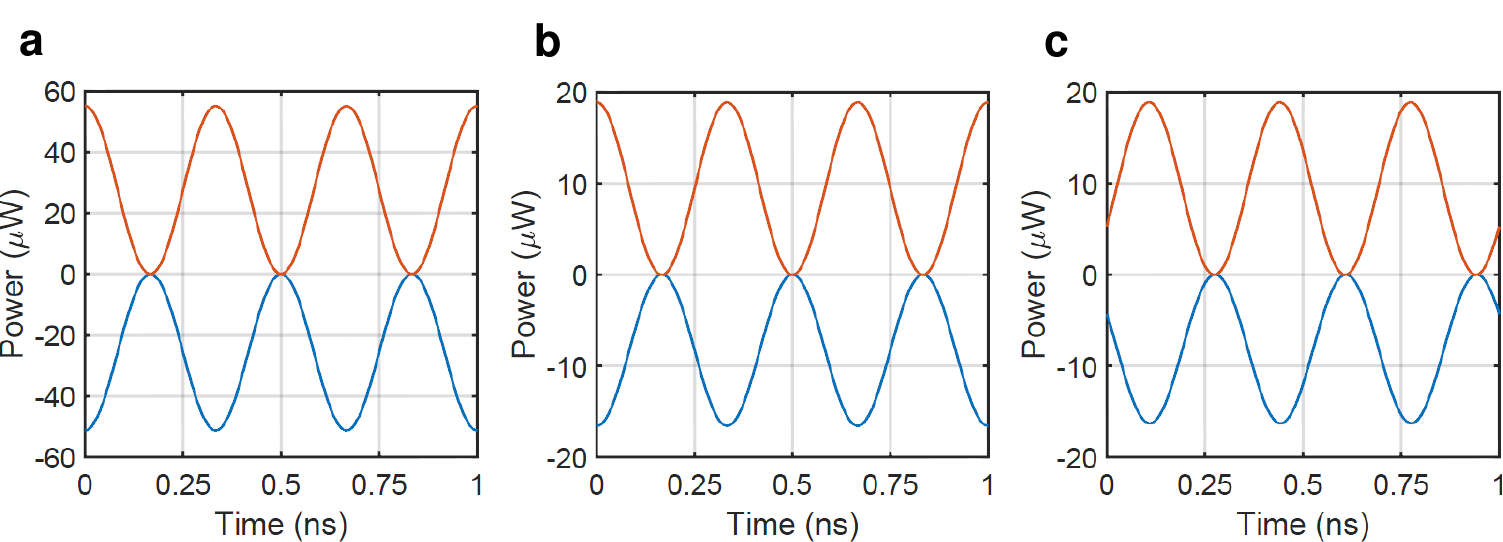}
\caption{{\bf Instantaneous power balance for a resonant dipole in receiving regime.} The cylindrical wire is illuminated by a normal incident electromagnetic plane wave. The length and the radius of the cylindrical wire is $l=93.9~\rm mm$ and $r=0.3~\rm mm$, respectively. {\bf a}, The wire is short circuited (zero-Ohm load). {\bf b}, the wire is loaded by a resistance $R_{\rm{load}}=50~\Omega$. {\bf c}, The wire is loaded by an inductance $L_{\rm{load}}=10~\rm nH$. In both {\bf b} and {\bf c}, the loads are lumped elements connected in a small gap at the center of the wire. Orange solid line always expresses the instantaneous radiated power and the blue solid line shows the summation of the reactive power, the dissipated power by the load resistance, and the power that is supplied to the dipole by the incident field.}
\label{fig_LD_RR_loaded}
\end{figure*}

%%%%%%%%%%%%%%%%%%%%%%%%%%%%%%%%%%%%%%%%%%%%%%%%%%%%%%%%%%%%%%%%%%%%%%%%%%%%%%%%%%%%%%%%%%%%%%%%%%

\section*{Numerical study}

In order to check the validity of our model, we do full-wave simulations (employing CST Microwave Studio) and assume time-harmonic excitation with no temporal modulation. We consider two different regimes: transmitting regime and receiving regime. In the first one, we connect an ideal source at the center of a short wire dipole in order to excite it. We numerically calculate the input power $P_{\rm{in}}(t)$, the reactive power $P_{\rm{reactive}}(t)$, and the radiated power  $P(t)$. Since we numerically have the electric current density, the dipole moment is readily found by integration of the current~\cite{Tretyakov1}.

Figure~\ref{fig_Pow_balance_dipoles_transmitting} shows the power balance for a resonant dipole. The length of the dipole is chosen so that the dipole resonates at about 1.5~GHz. Notice that by a simple scaling, the resonance can be shifted to other wavelengths. The reason to operate at microwave wavelengths is the fact that for simplicity we consider the dipole as a cylinder made of perfect electric conductor. The radius of the cylinder is supposed to be considerably smaller than the total length. As the figure shows, the instantaneous input power $P_{\rm in}(t)$ (blue color) and the radiated power $P(t)$ (red color) are equal in magnitude with a 180~degree phase difference. Therefore, the summation of the radiated power and the input power is zero. For the resonating dipole, the instantaneous reactive power vanishes since an equal exchange of electric and magnetic energy happens. An explicit analogy is drown to a resonant $LC$-circuit, where capacitance and inductance continuously exchange energy, keeping the total stored energy constant.

In the receiving regime, we excite the same resonant dipole by an incident electromagnetic plane wave. In this regime, the dipole is connected to a load instead of a source. Based on the load, three different cases are tested: The load is short circuited, the load is resistive, and the load is inductive. The results for instantaneous power balance are shown in Fig.~\ref{fig_LD_RR_loaded}. In the first case, the radiated power must be out of phase with the power which is equal to the negative of the  electromotive force $\mathcal{E}(t)$ multiplied by the electric current flowing through the load $i(t)$. In other words, $P(t)-\mathcal{E}(t)i(t)=0$. The electromotive force is analytically calculated since we know the incident electric field and the current distribution along the resonant dipole~\cite{Tretyakov1}. In the second case, in addition to the radiated power, we have dissipation due to the resistive load, and therefore we must add the corresponding active power. Thus, me have $P(t)+P_{\rm{resistance}}^{\rm{load}}(t)-\mathcal{E}(t)i(t)=0$. Finally, in the last case in which there is a reactive load, the instantaneous reactive power is not zero anymore and we should expect that $P(t)+P_{\rm{inductance}}^{\rm{load}}(t)-\mathcal{E}(t)i(t)=0$. As the figure illustrates, for all three cases, the summation of all powers is zero at all times confirming the validity of the instantaneous power balance approach. The important point is that the radiated power shown by orange color is exactly out of phase with the summation of other powers considered in blue color. 

\begin{figure*}[t!]\centering
\includegraphics[height=11cm]{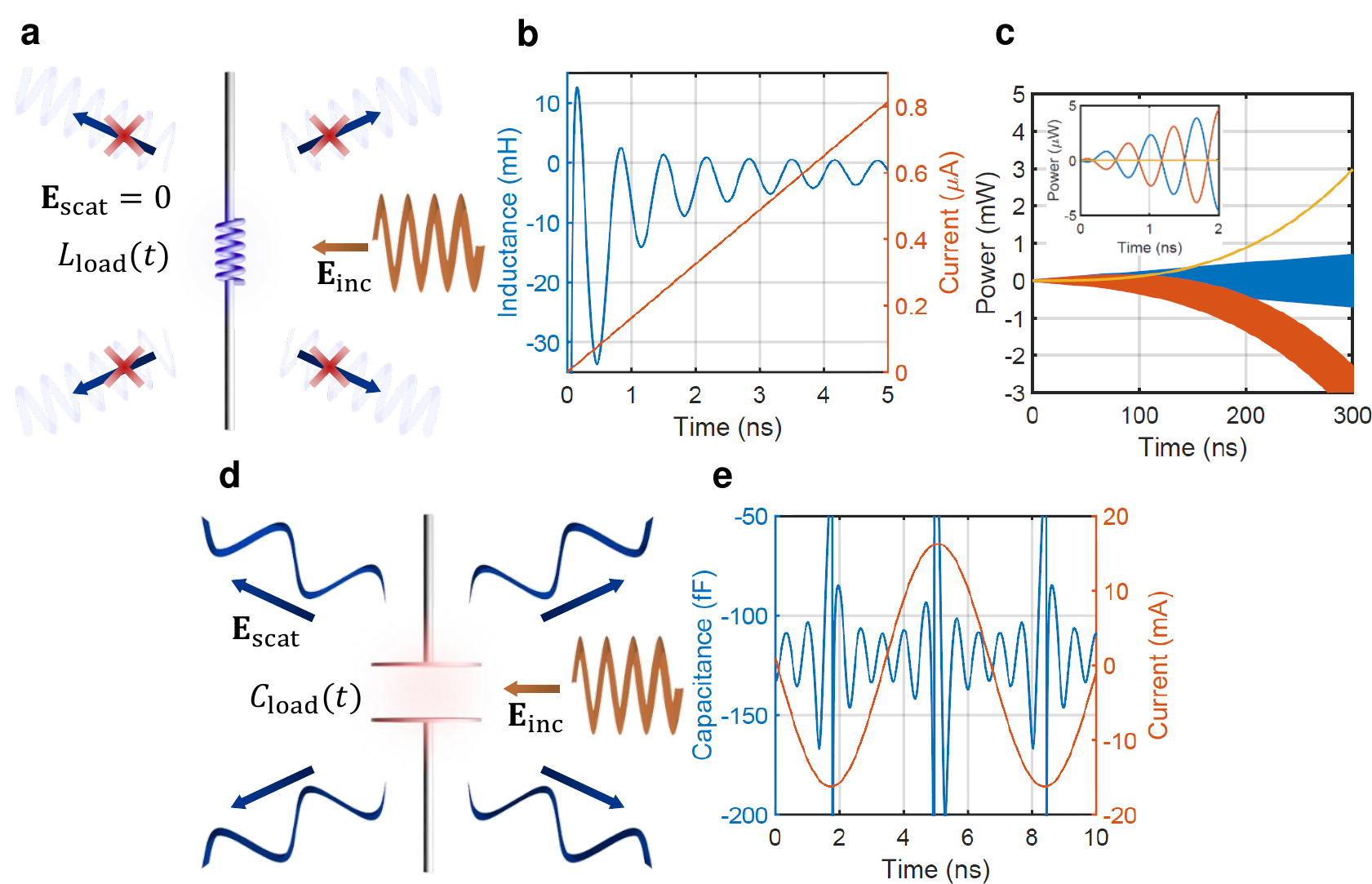}
\caption{{\bf Applications for meta-atoms modulated using time-varying reactive elements.} Electrically short dipole ($l=15~\rm mm$ and~$r=0.3~\rm mm$) is illuminated by a normally incident plane wave. {\bf a},~Schematic representation of a dipole loaded with a time varying inductance that cancels scattering. {\bf b},~Current through the inductance and function for the inductance that realize non-scattering regime shown in ({\bf a}). Note that the current that does not radiate is a linearly growing function.{\bf c},~Power balance in the case of linearly growing current. Reactive power grows proportionally to $t^3$ (orange line), amplitude of supplied power grows linearly (blue line) and the power produced by the time-varying inductance (red line) is such that it compensates both, supplied and reactive powers. The inset figure shows region near zero, when reactive power is negligible. {\bf d},~Schematic representation of a dipole loaded with a time varying capacitance that changes the frequency of the scattered wave. {\bf e},~Current through the capacitance and function for the capacitance that realize frequency shifting regime in ({\bf c}). The frequency of the scattered wave is shifted  from 1.5~GHz to 0.15~GHz.}
\label{fig_induc_capac_engineering_current}
\end{figure*}

%%%%%%%%%%%%%%%%%%%%%%%%%%%%%%%%%%%%%%%%%%%%%%%%%%%%%%%%%%%%%%%%%%%%%%%%%%%%%%%%%%%%%%%%%%%%%%%%%%%%%%%%

\section*{Applications}

Let us now discuss some examples of intriguing effects provided by time-modulated meta-atoms. The first example is about cancellation of dipole scattering. Regarding this case, we consider a small dipole antenna illuminated by a time-harmonic plane wave and select the time modulation of the load impedance so that the induced electric current is a linear function of time. Figure~\ref{fig_induc_capac_engineering_current}(b) shows the proper modulation function for a time-varying inductance as the load of the dipole. Recall  that the term responsible for  radiation is $\_{\dot{p}}(t)\cdot\_{\dot{\ddot{p}}}(t)$. Therefore, linearly growing or decaying current produces no radiation into the far zone: All the energy supplied to the dipole from the incident wave and via the time-modulated reactance is locked in the near field, although the dipole is located in free space. In the case of a linearly growing current, reactive energy stored in the antenna near field increases in time proportionally to $\sim t^3$ (orange line in Fig.~\ref{fig_induc_capac_engineering_current}(c)). Most of this energy is pumped into the system via the time-modulated inductance, with permanent energy exchange between the external field and the modulation source. The power supplied to the antenna from the incident wave is proportional to $\sim t\cos(\omega t)$~(blue line in Fig.~\ref{fig_induc_capac_engineering_current}(c)). Power supplied by the modulating force (red line in Fig.~\ref{fig_induc_capac_engineering_current}(c)) is such that it compensates both powers: reactive and incident. At the moment of time near zero, reactive power is negligibly small and almost does not participate in the power balance. It means that all the input energy is exchanged almost exclusively with the load. However, at later moments of time, reactive energy becomes rather significant.  
As long as the inductance is modulated according to this prescribed rule, the current in this receiving antenna as well as the energy stored in its near field will grow, without any loss to radiation. When the modulation will stop, this antenna will act as a usual transmitting antenna, radiating a high-amplitude pulse with the center frequency at the resonant frequency of the dipole. Conceptually, one can store reactive energy in such an open system without any limitation, increasing the modulation time. 

The second example is about shifting the wavelength of the excitation such that the scattering happens at a different wavelength. Figure~\ref{fig_induc_capac_engineering_current}(e) shows the required time-modulated capacitance of a small dipole antenna which allows us to shift the frequency from 1.5 GHz to 0.15 GHz, 10 times smaller. Obviously, the frequency of the scattered wave can be chosen arbitrary, therefore, such device can be named as any frequency multiplier or divider.

These are just two arbitrarily selected examples for showing the potential of dynamic meta-atoms for engineering their response.

%%%%%%%%%%%%%%%%%%%%%%%%%%%%%%%%%%%%%%%%%%%%%%%%%%%%%%%%%%%%%%%%%%%%%%%%%%%%%%%%%%%%%%%%%%%%%%%%%%%%%%%%

\section*{Conclusions} 

We have introduced a theoretical model for studying time-varying meta-atoms on the basis of the instantaneous power balance. We have derived the general time-domain equation for radiating dipole moments and, in particular, considered a time-modulated cylindrical wire antenna as an example. The validity of the proposed model was confirmed in two different regimes: Transmitting regime where the wire antenna is excited by an external lumped source, and receiving regime where the antenna is excited by an incident electromagnetic plane wave and loaded by time-variant lumped elements. Next, for the receiving regime, we have assumed temporally modulated reactive elements, such as time-varying   capacitance or inductance connected as a load to the cylindrical wire antenna. We have shown that properly choosing temporal modulations  of the reactive elements,  such dynamic scatterer can act as an invisible accumulator of energy or it can convert energy of an incident wave at the excitation frequency into scattered waves at an arbitarily chosen other frequency. This is our first step towards  understanding metamaterials/metasurfaces/metalines whose meta-atoms are time-variant. 
%Hence, experimental results confirming the theory and the numerical results written in this paper is reported in the near future. 
As the next step, it appears important to develop this model taking into account also magnetic response of dynamic meta-atoms. 

%%%%%%%%%%%%%%%%%%%%%%%%%%%%%%%%%%%%%%%%%%%%%%%%%%%%%%%%%%%%%%%%%%%%%%%%%%%%%%%%%%%%%%%%%%%%%%%%%%%%%%%%%%%%

\section*{Acknowledgements}
The authors thank F.~Liu and D.~Tzarouchis for having useful discussions.

\section*{Author contributions}
G.P. and M.S.M did the calculations and performed the numerical simulations under the supervision of S.A.T. All authors designed and conducted the research. Also, all authors contributed to the interpretation of the results and to writing the manuscript.

\section*{Competing interests}
The authors declare no competing interests.

\section*{Methods} 
All the numerical simulations are performed in CST Studio Suite. The transmitting regime of the dipoles was studied using a discrete port in the gap of the antenna. Current through the port and voltage over the port are extracted explicitly. In the receiving regime, in the gap of the antenna was placed a lumped element port with a negligibly small resistance, so that the current through the antenna center can be extracted. The voltage over the antenna gap is obtained using the electromotive force principle. Loaded dipoles in the receiving regime are modeled using lumped ports with specified values for $L$ and $C$. The dipole moment of the antenna is calculated integrating surface currents over the antenna surface.

\section*{Supplementary information} 

\subsection*{An $RC$ circuit with time-varying capacitance} 

\begin{figure}[h!]\centering
\includegraphics[width=0.3\textwidth]{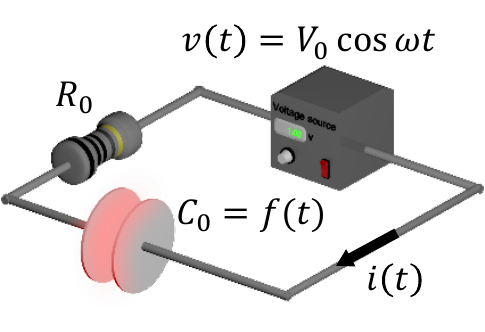}
\caption{Schematic view of an $RC$ circuit with time-modulated capacitance.} 
\label{fig:si}
\end{figure}

Let us assume a resistance $R_0$ and a time-varying capacitance $C_0(t)$ which are connected to a voltage source $v(t)$ in series, as shown in Fig.~\ref{fig:si}. The current $i(t)$ flowing through the capacitance and the voltage $v_{\rm{C}_0}(t)$ over the capacitance are related as
\begin{equation}
i(t)={d\over dt}\bigg[C_0(t)v_{\rm{C}_0}(t)\bigg].
\end{equation}
Since we can write 
\begin{equation}
v(t)=v_{\rm{R}_0}(t)+v_{\rm{C}_0}(t),
\end{equation}
the electric current can be readily found via the following first-order differential equation: 
\begin{equation}
{d i(t)\over dt}+a(t)i(t)=b(t),
\end{equation}
in which
\begin{equation}
\begin{split}
a(t)=&{\displaystyle 1+R_0{dC_0(t)\over dt}\over R_0C_0(t)},\cr
b(t)=&{\displaystyle {dC_0(t)\over dt}v(t)+C_0(t){dv(t)\over dt}\over R_0C_0(t)}.
\end{split}
\end{equation}
Coefficients $a(t)$ and $b(t)$ depend on the time-varying capacitance and its time derivative. It means that by proper choosing  $C_0(t)$ we have a possibility to manipulate the electric current in the desired fashion. We can rewrite the above differential equation  in order to determine the required capacitance for realizing the desired electric current. After some algebraic manipulations, we find 
\begin{equation}
{d C_0(t)\over dt}+m(t)C_0(t)=n(t),
\end{equation}
where 
\begin{equation}
\begin{split}
m(t)=&{\displaystyle {dv(t)\over dt}-R_0{di(t)\over dt}\over v(t)-R_0i(t)},\cr
b(t)=&{\displaystyle {i(t)\over v(t)-R_0i(t)}}.
\end{split}
\end{equation}
Recall that the general solution of the above differential equation can be expressed as  
\begin{equation}
C_0(t)=\exp\big(-\int m(t)dt\big)\Bigg[\int n(t)\exp\big(\int m(t)dt\big)dt+Y\Bigg],
\end{equation}
in which $Y$ depends on the initial condition.

%%%%%%%%%%%%%%%%%%%%%%%%%%%%%%%%%%%%%%%%%%%%%%%%%%%%%%%%%%%%%%%%%%%%%%%%%%%%%%%%%%%%%%%%%%%%%%%%%%%%%%%%%%

\subsection*{Instantaneous radiated power for a Hertzian dipole from the induced electromotive force perspective}

Here, we provide an alternative derivation for the instantaneous radiated power based on the induced electromotive force method. 
Let us consider an oscillating Hertzian dipole with the current density
\begin{equation}
\_J=Il\delta(\_r)~\_a_z. 
\end{equation} 
The dipole has the length $l$ and carries the electric current whose amplitude is $I$ (here we consider the time-harmonic regime: $i(t)=I\cos(\omega t)$). The dipole is oriented along axis $z$ ($\_a_z$ is the unit vector along the $z$-axis), and located at the origin of the spherical coordinate system $r=0$. Using the induced electromotive force method, we find the instantaneous radiated power from such Hertzian dipole. We know that the theta component ($\_a_\theta$) of the electric field generated by the Hertzian dipole is expressed as 
\begin{equation}
\_E=j\omega\mu_0{{Il}\over4\pi}\Bigg[1+{c\over j\omega r}-{c^2\over\omega^2r^2}\Bigg]{e^{\big({-j\omega r\over c}\big)}\over r}\sin\theta~\_a_\theta,
\label{eq:fieldtheta}
\end{equation}
where $\omega$ is the angular frequency, $\mu_0$ and $\epsilon_0$ represent the permeability and permittivity of vacuum, respectively, and $c$ denotes the speed of light. We are interested in examining the electric field at very small distances from the dipole: $r\rightarrow 0$.
While the total electric field is singular at the source location, there is a non-singular component. It can be found expanding the exponential function in the above equation:
\begin{equation}
e^{-j\omega r/c}\approx 1-j{\omega r\over c}-{\omega^2r^2\over2c^2}+j{\omega^3r^3\over 6c^3}. 
\label{eq:exponfield}
\end{equation}
Keeping these four terms in the expansion is sufficient to find the non-singular part.  To do that, we substitute Eq.~(\ref{eq:exponfield}) into Eq.~(\ref{eq:fieldtheta}) and obtain     
\begin{equation}
\_E{_{\rm{ns}}}={(j\omega)^2\mu_0\over6\pi c}Il~\_a_z.
\label{eq:jomegafield}
\end{equation}
At this point, we can use the inverse Fourier transform to return to the time domain. In Eq.~(\ref{eq:jomegafield}), $(j\omega)^2$ corresponds to the second time derivative of the electric current. Thus,
\begin{equation}
\_E{_{\rm{ns}}}(t)={\mu_0\over6\pi c}{d^2i(t)\over dt^2}l~\_a_z.
\end{equation}
Knowing the time-domain expression of the non-singular component of the electric field parallel to the dipole moment, we obtain the corresponding electromotive force (assuming that the dipole size is negligibly small as compared with all relevant wavelengths): 
\begin{equation}
\mathcal{E}(t)=-\int_l\_E{_{\rm{ns}}}(t)\cdot\_{dl}=-{\mu_0\over6\pi c}{d^2i(t)\over dt^2}l^2.
\end{equation}
Expressing  the electric current in terms of the  corresponding electric dipole moment,  
\begin{equation}
{d^3p(t)\over dt^3}=l{d^2i(t)\over dt^2},
\end{equation}
the formula for the electromotive force reduces to 
\begin{equation}
\mathcal{E}(t)=-{\mu_0\over6\pi c}{d^3p(t)\over dt^3}l.
\label{eq:elmftim}
\end{equation}
After some simple algebraic manipulations and recalling  that $i(t)=(1/l)~dp(t)/dt$, finally, we derive the instantaneous radiated power radiated from the Hertzian dipole: 
\begin{equation}
P_{\rm{rad}}(t)=\mathcal{E}(t)i(t)=-{\mu_0\over6\pi c}{d\_p(t)\over dt}\cdot{d^3\_p(t)\over dt^3},
\end{equation} 
which is the same as \eqref{Power}. Note that the direct inverse Fourier transform of the usual expression for power radiated from a time-harmonic dipole source gives a different result, where the oscillating part of the power is missing.

In the following subsection, we present a general derivation considering any dipolar source of radiation, and we prove that the above equation holds for any kind of current distribution. However, before that, we are interested in elucidating what we derived for the Hertzian dipole. The radiation-reaction force associated with a non-relativistic electron as it accelerates is 
\begin{equation}
\_F_{\rm{reaction}}(t)={\mu_0\over6\pi c}q^2{d\_a\over dt},
\end{equation}
where $\_a$ denotes the acceleration and $q$ represents the electron charge. Since the acceleration is the second time derivative of the position vector,  the radiation-reaction force is proportional to the third time derivative of the dipole moment:  
\begin{equation}
\_F_{\rm{reaction}}(t)={\mu_0\over6\pi c}{d^3\_p (t)\over dt^3}q.
\end{equation}
If we compare this force with what is written in Eq.~(\ref{eq:elmftim}), we see that the concept of the induced electromotive force is quite similar to that of the radiation-reaction force. We can also perceive this fact from the way how we derive the radiated power. To find  the total power radiated  by the accelerated electron, we  calculate $P_{\rm{electron}}=-\_F_{\rm{reaction}}(t)\cdot\_v(t)$ where $\_v$ is the velocity of the electron. Notice that $q\_v(t)=d\_p(t)/dt$. Thus, we have
\begin{equation}
P_{\rm{electron}}=-{\mu_0\over6\pi c}{d\_p(t)\over dt}\cdot{d^3\_p(t)\over dt^3},
\end{equation}
which is exactly the same expression as for the  power radiated  from the Hertzian dipole. To obtain the total power radiated  by the Hertzian dipole, we have multiplied the electromotive force by the electric current: $P_{\rm{Hertzian~Dipole}}=\mathcal{E}(t)i(t)$, in which the electric current is related to the charge velocity.

%%%%%%%%%%%%%%%%%%%%%%%%%%%%%%%%%%%%%%%%%%%%%%%%%%%%%%%%%%%%%%%%%%%%%%%%%%%%%%%%%%%%%%%%%%%%%%%%%%%%%%%%%%%%

\subsection*{Instantaneous radiated power for an electric dipole moment from a general perspective}

The charge and current densities are expanded into power series of $|\_r-\_r^\prime|/c$, assuming that $|\_r-\_r^\prime|/c$ is small compared to the characteristic time of variation of the charged-particle system (non-relativistic approximation). Under this assumption, the expansion for the potentials results in 
\begin{equation}
\begin{split}
&\phi(\_r ,t)\approx\frac{1}{4\pi\epsilon_0}\Bigg[
\int_{V_{\rm{c}}}\frac{\rho(\_r^\prime ,t)}{|\_r-\_r^\prime|}d\_r^\prime
-\int_{V_{\rm{c}}}\frac{\dot{\rho}(\_r^\prime,t)}{c}d\_r^\prime+\cr
&\int_{V_{\rm{c}}}\frac{\ddot{\rho}(\_r^\prime,t)}{2c^2}|\_r-\_r^\prime|d\_r^\prime-
\int_{V_{\rm{c}}}\frac{\dot{\ddot{\rho}}(\_r^\prime ,t)}{6c^3}|\_r-\_r^\prime|^2d\_r^\prime+...\Bigg].
\end{split}
\end{equation} 
and 
\begin{equation}
A(\_r ,t)\approx\frac{\mu_0}{4\pi}\Bigg[\int_{V_{\rm{c}}}\frac{\_J(\_r^\prime,t)}{|\_r-\_r^\prime|}d\_r^\prime-\int_{V_{\rm{c}}}{{\dot{\_J}}(\_r^\prime,t)\over c}d\_r^\prime+...\Bigg].
\label{eq:MAE}
\end{equation}
Let us start from the scalar potential and consider the first four terms in the expansion. The first term is the static contribution and accordingly we can write that 
\begin{equation}
\int_{V_{\rm{c}}}\_J\cdot\nabla\phi\,\vert_{\rm{term\,1}}={d\over dt}\Bigg[{1\over8\pi\epsilon_0}\int_{V_{\rm{c}}}{\rho(\_r,t)\rho(\_r^\prime,t)\over|\_r-\_r^\prime|}d\_rd\_r^\prime \Bigg].
\label{eq:ter1jphi}
\end{equation}
Notice that we use the equation of continuity:
\begin{equation}
\nabla\cdot\_J=-{\partial\rho\over\partial t},
\end{equation}
for deriving expression  Eq.~(\ref{eq:ter1jphi}). The second term in the expansion does not contribute to the energy transfer since for the dipole radiation we have
\begin{equation}
{1\over c}\int_{V_{\rm{c}}}{\partial\rho(\_r^\prime,t)\over\partial t}d\_r^\prime=0.
\end{equation}
The third term, which is related to the second derivative of the charge density, is certainly a dynamic term, and we can calculate the contribution of this term into the energy transfer in a similar way as we do for the first term. Hence, 
\begin{equation}
\begin{split}
&\int_{V_{\rm{c}}}\_J\cdot\nabla\phi\,\vert_{\rm{term\,3}}=\cr
&{d\over dt}\Bigg[{\mu_0\over16\pi}\int_{V_{\rm{c}}}{\partial\rho(\_r,t)\over\partial t}{\partial\rho(\_r^\prime,t)\over\partial t}|\_r-\_r^\prime|d\_rd\_r^\prime \Bigg].
\end{split}
\end{equation}
The fourth term is important since it is associated with the dipole radiation. The gradient of this term is the dynamic force exerted on one unit charge: 
\begin{equation}
\nabla\Bigg[\int_{V_{\rm{c}}}\frac{\dot{\ddot{\rho}}(\_r^\prime ,t)}{6c^3}|\_r-\_r^\prime|^2d\_r^\prime\Bigg]={1\over 3c^3}\int_{V_{\rm{c}}}|\_r-\_r^\prime|\dot{\ddot{\rho}}(\_r^\prime ,t)d\_r^\prime.
\label{eq:M1}
\end{equation}
Since the dipole moment is defined as 
\begin{equation}
\_p=\int_{V_{\rm{c}}}\_r^\prime\rho(\_r^\prime,t)\,d\_r^\prime,
\end{equation}
Eq.~(\ref{eq:M1}) is simplified and reduces to
\begin{equation}
\nabla\Bigg[-\int_{V_{\rm{c}}}\frac{\dot{\ddot{\rho}}(\_r^\prime ,t)}{6c^3}|\_r-\_r^\prime|^2d\_r^\prime\Bigg]={1\over3c^3}\dot{\ddot{\_p}}(t).
\end{equation}
Having the time derivative of the dipole moment:
\begin{equation}
{d\_p\over dt}=\int_{V_{\rm{c}}}\_J\,d\_r,
\end{equation}
we can finally write 
\begin{equation}
\int_{V_{\rm{c}}}\_J\cdot\nabla\phi\,\vert_{\rm{term\,4}}={\mu_0\over12\pi c}\dot{\_p}(t)\cdot\dot{\ddot{\_p}}(t).
\end{equation}
Next is the vector potential whose study appears easier. The first term in the expansion in Eq.~(\ref{eq:MAE}) corresponds to the static part and therefore we can readily conclude that 
\begin{equation}
\int_{V_{\rm{c}}}\_J\cdot{\partial\_A\over\partial t}\vert_{\rm{term\,1}}={d\over dt}\Bigg[{\mu_0\over8\pi}\int_{V_{\rm{c}}}\frac{\_J(\_r^\prime,t)\cdot\_J(\_r,t)}{|\_r-\_r^\prime|}d\_r^\prime d\_r\Bigg].
\end{equation}
However, similar to the fourth term related to the scalar potential, here the second term is associated with the radiated power. The time derivative of the vector potential is the second time derivative of the second term in expansion which indeed means the third time derivative of the dipole moment. Therefore, 
\begin{equation}
\int_{V_{\rm{c}}}\_J\cdot{\partial\_A\over\partial t}\vert_{\rm{term\,2}}=-{\mu_0\over4\pi c}\dot{\_p}(t)\cdot\dot{\ddot{\_p}}(t).
\end{equation}
From the above equations, the transferred energy can be obtained as 
\begin{equation}
\begin{split}
&-\int_{V_{\rm{c}}}\_J\cdot\_E\,d\_r=-\int_{V_{\rm{c}}}\_J\cdot\bigg(-\nabla\phi-{\partial\_A\over\partial t}\bigg)\,d\_r\approx\cr&-\frac{\mu_0}{6\pi c}\_{\dot{p}}(t)\cdot\_{\dot{\ddot{p}}}(t)+
\frac{d}{dt}\Bigg[\frac{1}{8\pi\epsilon_0}\int_{V_{\rm{c}}}\frac{\rho(\_r^\prime,t)\rho(\_r,t)}{|\_r-\_r^\prime|}d\_r^\prime d\_r\cr
&+\frac{\mu_0}{16\pi}\int_{V_{\rm{c}}}\frac{\partial\rho(\_r,t)}{\partial t}\frac{\partial\rho(\_r^\prime,t)}{\partial t}|\_r-\_r^\prime|\,d\_r\,d\_r^\prime\cr
&+\frac{\mu_0}{8\pi}\int_{V_{\rm{c}}}\frac{\_J(\_r^\prime,t)\cdot\_J(\_r,t)}{|\_r-\_r^\prime|}\,d\_r^\prime d\_r\Bigg],
\end{split}
\end{equation} 
which is \eqref{eq_energy_transfer_all_terms} of the main text.

\end{document}